\documentclass[10pt]{article}
\usepackage{sao2}
\usepackage{psfig,epsf}

\issue{2008, 63, 389--396}

\def\apj{Astrophys. J}

\def\apjs{Astrophys. J. Supp.}
\def\ijmpd{Int. J. Mod. Phys. D}

\def\prd{Phys.\ Rev.\ D}

\def\wmap{\hbox{\sl WMAP~}}
\def\glesp{G{\sc lesp }}
\def\sqd{\hbox{\rlap{$\sqcap$}$\sqcup$}$^\circ$}

\begin{document}
\markboth{Verkhodanov, Khabibullina, Majorova, Parijskij}
    {Correlation properties of the maps of the NVSS survey and WMAP ILC}
\title{Correlation properties of the maps of the NVSS survey and WMAP ILC}
\author{
O.V.~Verkhodanov\inst{a}
\and M.L.~Khabibullina\inst{a}
\and E.K.~Majorova\inst{a}
\and Yu.N.~Parijskij\inst{a}
}
\institute{
\saoname
}
\date{June 9, 2008}{June 30, 2008}
\maketitle

\begin{abstract}
In this paper we study one-dimensional sections of the maps of WMAP ILC and
of the  NVSS survey on scale lengths of 0.75, 3, 4.5, and 6.75 degrees and
analyze the correlation properties of the sections.
On these maps we identify the domains where the absolute value of
the correlation coefficient exceeds 0.5.
The catalog of such domains is presented. It is shown that the number of
the domains agrees with the number of
such domains on simulated maps and this fact may be indicative of just
statistical agreement of the
arrangement of the domains considered.
%
\\
\mbox{\hspace{1cm}}
\end{abstract}

\section{INTRODUCTION}

With the advent of high-precision millimiter-, centimeter-, and
decimeter-wave radio-astronomical surveys
it became possible to apply correlation studies to test the
self-consistentcy of the cosmological models.
The opening of public access to one-, three-, and five-year accumulated
data on the measurements of cosmic microwave
background (CMB) on the entire sky
sphere~\cite{wmapresults:Verkhodanov_n,wmapfg:Verkhodanov_n,
wmappara:Verkhodanov_n,wmap3ytem:Verkhodanov_n,wmap3ycos:Verkhodanov_n,%
wmap5ytem:Verkhodanov_n,wmap5ycos:Verkhodanov_n}
conducted using  WMAP\footnote{\tt http://lambda.gsfc.nasa.gov}
(Wilkinson Microwave
Anisotropy Probe) became a real breakthrough in this field.
To reconstruct the CMB signal from multifrequency
observations, it is used the method of inner linear combinations
(ILC---Internal Linear Combination) of
background components~\cite{wmapresults:Verkhodanov_n},
which yielded a map of microwave background, also
known as the ILC map, which is used to analyze low harmonics with multipole
numbers  $\ell\le100$. The
ILC was constructed from observations made in five channels: 23\,GHz
(the K band), 33\,GHz (the Ka band),
41\,GHz (the Q  band), 61\,GHz (the V band), and 94\,GHz (the W band).

Quite a few papers were devoted to analyzing the statistical properties
of the signal in the reported ILC map and
discussing the non-Gaussian nature of the data of this map revealed using
various methods including phase
analysis~\cite{nong:Verkhodanov_n,ndv03:Verkhodanov_n,ndv04:Verkhodanov_n,%
coles04:Verkhodanov_n}, Maxwellian
multipole vectors~\cite{copi:Verkhodanov_n,copi3y:Verkhodanov_n},
wavelets~\cite{vielva04:Verkhodanov_n,mw_wavel:Verkhodanov_n,%
cruz05:Verkhodanov_n,cruz06:Verkhodanov_n}, and
\mbox{Minkowski} functionals~\cite{eriksenmf:Verkhodanov_n,%
park3y:Verkhodanov_n}.

Non-Gaussian features extensively discussed by the researchers include
the so-called Cold Spot (hereafter
referred to as CS) with the size of about $10\degr$
(Galactic coordinates  $l=209\degr$, \mbox{$b=-57\degr$})
found in the distribution of cosmic microwave background in
the Southern Hemisphere. The CS, which is statistically
identified in the distribution of CMB, is one of the features of
the map that are inconsistent with the
hypothesis of uniform Gaussian background fluctuations. It was
initially pointed out as a feature that deviates
from Gaussian statistics when applying \mbox{wavelet analysis
\cite{vielva04:Verkhodanov_n,cayon05:Verkhodanov_n,cruz05:Verkhodanov_n,%
cruz06:Verkhodanov_n}}
to the data of the first and third years of observations of the  WMAP mission.

A decrease of the space density of radio sources was found later in the
NVSS maps
\cite{rudnick:Verkhodanov_n}, and this discovery led the researches
to conclude that there must exist a giant
140-Mpc-large void at a redshift of  \mbox{$z<1$,} which produces
the gravitational anomaly, which
causes the integrated Sachs--Wolfe effect~\cite{swe:Verkhodanov_n}
and shows up as the  CS.
The researchers also discussed the possibility of explaining the spot
in terms of the Sunyaev--Zel'dovich
effect~\cite{sz:Verkhodanov_n} due to the scattering in the Eridanus
supergroup of galaxies~
\cite{cruz05:Verkhodanov_n,brough_eridan:Verkhodanov_n}.

Exotic interpretations of the effect included the texture---a topological
defect during the phase
transition in the early Universe~\cite{texture:Verkhodanov_n,%
texture2:Verkhodanov_n} and Bianchi VII$_h$
anisotropic cosmological model~\cite{jaffebianchi:Verkhodanov_n}.

At the same time, serious arguments were advanced suggesting the map
contains  residual contribution
of Galactic background components, which produces the observed deviation
from the Gaussianity.
This contribution can show up in the form of the earlier discovered
relation between the cleaned
microwave background map and Galactic components of radiation
\cite{pecquad:Verkhodanov_n,nv_quad:Verkhodanov_n} and may also
specity the properties of low-order
multipoles \mbox{$\ell\le20$,} resulting in their unstable reconstruction
\cite{instab:Verkhodanov_n}. 
In particular, these multipoles may explain the specific features
of the CS, which demonstrate the deviation of the statistics of
clusters of signal oscillation peaks
in the vicinity of the spot~\cite{nas_cs:Verkhodanov_n}---the increase
of the number of positive peaks.

An independent analysis of the spot's properties found in the region
with close coordinates on maps and
in radio-source counts in the NVSS survey~\cite{nvss:Verkhodanov_n} showed
that the Cold Spot whose giant size and existence are difficult to explain
in terms of the cosmological $\Lambda$CDM-model may be a common statistical
deviation due to systematic effects
\cite{no_cs:Verkhodanov_n}.

In this paper we decided to verify the correlation properties of
the NVSS and ILC maps on angular scales
of 0.75, 3, 4.75, 6.75, and 9.75 degrees and to try to find and analyze
CS type spots. We chose the angular
scales to be much greater than the size of the smoothing window of the
ILC map ($1\degr$), but smaller than
$10\degr$ in order to preserve a sufficient number of pixels for statistical
analysis. Our statistical analysis
also included a sub-degree angular scale \mbox{($45\arcmin=0.75\degr$).}
In this paper we use the method of cross-correlation within selected
sections proposed by
\mbox{Khabibullina~et~al.~\cite{rzf_cmb:Verkhodanov_n,%
rzf_cmb2:Verkhodanov_n}} and employed to analyze low-frequency spatial
fluctuations of microwave background in the domain covered by the RZF
survey performed with the RATAN-600
radio telescope~\mbox{\cite{rzf:Verkhodanov_n,rzf_src:Verkhodanov_n}}.
To assess the results obtained, we simulated the
maps by generating random Gaussian fluctuations with the power spectrum
as implied by the $\Lambda$CDM cosmological
model.

\section{THE DATA}

The ILC map bears the data on microwave background with a resolution of
up to  $\ell\le100$ \mbox{(about $1\degr$).}
This map was reconstructed based on multifrequency observations in
five different wavelength intervals made
over the entire celestial sphere and available from the web-site of
the WMAP mission.

The NRAO VLA Sky Survey (NVSS) \cite{nvss:Verkhodanov_n} is currently
the most complete high-sensitivity and
large-area survey in the Northern sky. It was performed with a VLA
at a frequency of  1.4\,GHz from 1993
until 1996 and covers the entire Northern sky above the declination
of $\delta=-40^\circ$ (33884\sqd or
82\% of the sly sphere). The survey is extensively used for various
statistical cosmological studies. The
catalog contains a total of $1.8\times10^6$ sources and, according
to its description, is 99\% and 50\% complete
down to the flux densities of $S_{1.4\,\mbox{\tiny{\rm GHz}}}=3.5$\,mJy
and 2.5\,mJy, respectively. The survey was
made in the D-configuration of the VLA and the half-power size of
the synthesized beam, which determines the
resolution, was equal to about 45 arcsec. The data obtained as a result of
the NVSS survey are available at the
NRAO web-site\footnote{\tt http://www.cv.nrao.edu/nvss/},
via the SkyView virtual
telescope~\footnote{\tt http://skyview.gsfc.nasa.gov}, and from the CATS
database~\footnote{\tt http://cats.sao.ru}
\cite{cats:Verkhodanov_n,cats2:Verkhodanov_n}.

\subsection{Properties of the NVSS catalog and the maps}
We analyzed the maps of the NVSS survey by constructing one-dimensional
sections with the
smoothing due to the beam as described
by Majorova~\cite{nvss_rzf:Verkhodanov_n}. Let us now consider two types
of NVSS data: (1) the initial maps of the survey and (2) the catalog of
radio sources. In second case, we
also set the threshold for elimination of sources from the scans as
a free parameter, thereby using the possibility
of forming scans that are free of the influence of powerful radio sources.

To compare the sections constructed by different methods, we compute
the correlation coefficients for: (1)
scans based on the catalog and maps and \linebreak (2) scans computed
for two-dimensional maps and one-dimensional sections
running through the center of the selected area.

In first case the correlation coefficient between all the sections
exceeds 0.99, i.e., the sections computed
based on catalog data and those based on the maps can be viewed as
identical for our purposes, making our work much
easier and allowing the source lists from the catalog to be used for
the analysis. In latter case the correlation
coefficient was 0.85 or higher, which we consider to be satisfactory
and acceptable for fast computations and for the
corresponding analysis.

To analyze the statistical correlation properties, we transformed
the NVSS maps into a set of combined
one-dimensional sections ($M$-scans) at given declinations with
right ascensions ranging from  0$^h$ to 24$^h$.
We chose the declination step between the sections to be equal
to the size of the chosen pixel on the sphere. The
values in each pixel in modified one-dimensional NVSS sections are
equal to the mean-squared source fluxes within
the area of given size centered on the given pixel:
\begin{eqnarray}
    M_i = \frac{1}{N_r} \sum_{\kappa_i}^{N_r} S_{\kappa_i}^2, \quad
    P_{\kappa_i}(r) \in  R_i \,,
\end{eqnarray}
\noindent
where $N_r$ is the number of sources in the pixel (elementary sky area)
of the given size of the NVSS survey;
$R_i$ is the pixel number $i$ of given size; $S_{\kappa_i}$ is the flux
of the radio source number
$\kappa_i$ in pixel $R_i$, and $P_{\kappa_i}(r)$ is the position of
source number $\kappa_i$.
The pixel size $R_i$ is specified by the correlation scale 0.75, 3, 4.5,
6.75, and 9.75$\degr$. Such a characterization of the  NVSS map makes
it possible to describe the distribution
of sources in the given sky area, which varies from pixel to pixel and,
in particular, allows statistical
verification of the presence of hidden fluctuations due to
the incompleteness of the  \mbox{WMAP catalog
\cite{wmapfg:Verkhodanov_n,wmap_src5:Verkhodanov_n,trushkin:Verkhodanov_n}.}
We exclude from our analysis the area in the NVSS survey at
declinations $\delta<-40\degr$ and the sections that
cross it.

\begin{table*}[!tbp]
\begin{center}
\caption{Coordinates of the centers of the areas with the correlation
coefficients $|{\rm K}|\ge0.5$. The coordinates:
right ascension $\alpha$ in hours and declination $\delta$ in degrees
(for the epoch of 2000.0) are accurate
to about half the pixel size. The size of the pixel side is $0.75\degr$
}
\begin{tabular}{r|r|r||r|r|r||r|r|r}
\hline
$\alpha^h$&$\delta^\circ$&$k$&$\alpha^h$&$\delta^\circ$&$k$&$\alpha^h$&$\delta^\circ$&$k$\\
\hline
14.407 & -32.87  &  -0.58 & 8.387  &  12.12  &   0.52 & 6.024  &  45.12 &   -0.55 \\
12.332 & -37.37  &  -0.52 & 21.711 &  13.62  &  -0.52 & 7.228  &  45.12 &   -0.62 \\
12.647 & -37.37  &  -0.52 & 3.431  &  15.88  &   0.52 & 7.413  &  52.62 &   -0.50 \\
12.961 & -37.37  &  -0.59 & 3.535  &  15.88  &   0.51 & 9.472  &  52.62 &   -0.57 \\
10.002 & -32.87  &  -0.50 & 3.639  &  15.88  &   0.55 & 18.203 &  52.62 &   -0.59 \\
6.747  & -29.87  &   0.64 & 3.795  &  15.88  &   0.51 & 18.054 &  57.12 &   -0.54 \\
11.517 & -16.37  &  -0.54 & 7.342  &  30.12  &  -0.56 & 23.120 &  57.12 &   -0.50 \\
11.726 & -16.37  &  -0.56 & 17.805 &  30.12  &  -0.52 & 22.379 &  59.38 &    0.50 \\
18.969 & -16.37  &  -0.57 & 17.920 &  30.12  &  -0.65 & 1.104  &  60.12 &    0.55 \\
7.574  & -10.37  &  -0.52 & 22.719 &  30.12  &  -0.51 & 5.931  &  62.38 &   -0.51 \\
7.726  & -10.37  &  -0.50 & 0.716  &  33.12  &  -0.52 & 6.245  &  63.88 &    0.52 \\
0.700  & -1.38   &   0.51 & 11.001 &  36.88  &   0.59 & 8.454  &  71.38 &   -0.57 \\
3.751  & -1.38   &   0.56 & 14.313 &  36.88  &   0.51 & 9.550  &  71.38 &   -0.55 \\
5.402  & -1.38   &  -0.51 & 14.938 &  36.88  &   0.53 & 9.863  &  71.38 &   -0.51 \\
5.702  & -1.38   &  -0.51 & 3.508  &  38.38  &  -0.51 & 12.698 &  78.88 &    0.53 \\
2.313  &  6.12   &   0.59 & 19.006 &  38.38  &  -0.55 & 13.734 &  78.88 &    0.64 \\
3.938  & 12.12   &   0.51 & 4.864  &  41.38  &  -0.50 & 1.169  &  82.62 &   -0.62 \\
8.080  & 12.12   &   0.52 & 5.464  &  41.38  &  -0.52 &        &        &         \\
\hline
\end{tabular}
\end{center}
\end{table*}

\begin{table*}[!htbp]
\begin{center}
\caption{Coordinates of the centers of the areas with the correlation
coefficients $|{\rm K}|\ge0.5$. The coordinates:
right ascension $\alpha$ in hours and declination $\delta$ in degrees
(for the epoch of 2000.0) are accurate
to about half the pixel size. The size of the pixel side is $3\degr$
}
\begin{tabular}{r|r|r||r|r|r||r|r|r}
\hline
$\alpha^h$&$\delta^\circ$&$k$&$\alpha^h$&$\delta^\circ$&$k$&$\alpha^h$&$\delta^\circ$&$k$\\
\hline
23.448 &-35.88 &  0.57  &   3.451  &  9.88 &  -0.54 &    8.207  & 25.62 &  -0.63  \\
0.807  & -7.38 &  -0.60 &   12.586 &  9.88 &  -0.51 &    9.538  & 25.62 &  -0.51  \\
11.293 & -7.38 &  -0.57 &   2.271  & 14.38 &  0.51  &    22.625 & 25.62 &  -0.54  \\
14.117 & -7.38 &  -0.63 &   16.724 & 14.38 &  0.63  &    3.931  & 30.12 &  -0.56  \\
16.133 & -7.38 &  -0.56 &   20.027 & 14.38 &  0.53  &    15.724 & 30.12 &  -0.53  \\
23.595 & -7.38 &  -0.52 &   21.885 & 14.38 &  0.55  &    21.871 & 40.62 &  -0.61  \\
6.820  & -4.38 &  0.53  &   3.397  & 19.62 &  -0.55 &    1.984  & 45.12 &  0.53   \\
21.061 & -4.38 &  0.51  &   6.158  & 19.62 &  -0.55 &    4.535  & 45.12 &  0.57   \\
8.206  & -2.12 &  0.51  &   8.281  & 19.62 &  -0.51 &    5.953  & 45.12 &  0.52   \\
0.800  & -1.38 &  0.53  &   2.787  & 21.12 &  -0.51 &    2.965  & 66.12 &  0.52   \\
 0.800 & -0.62 &  0.52  &   3.645  & 21.12 &  -0.59 &           &       &         \\
19.400 &  0.12 &  0.55  &   8.405  & 21.88 &  -0.66 &           &       &         \\
\hline
\end{tabular}
\end{center}
\end{table*}

\begin{table*}[tbp]
\begin{center}
\caption{Coordinates of the centers of the areas with the correlation
coefficients $|{\rm K}|\ge0.5$. The coordinates:
right ascension $\alpha$ in hours and declination $\delta$ in degrees
(for the epoch of 2000.0) are accurate
to about half the pixel size. The size of the pixel side is $4.5\degr$
}
\begin{tabular}{r|r|r||r|r|r||r|r|r}
\hline
$\alpha^h$&$\delta^\circ$&$k$&$\alpha^h$&$\delta^\circ$&$k$&$\alpha^h$&$\delta^\circ$&$k$\\
\hline
21.608 &  -32.12&   0.52 &  15.218 &  18.88 &  -0.52 &  22.444 &  35.38 &  -0.51 \\
12.300 &  0.12  & -0.50  &  16.925 &  35.38 &  -0.50 &  23.213 &  57.12 &  0.57  \\
22.500 &  0.12  & 0.55   &  18.764 &  35.38 &  -0.55 &         &        &        \\
13.950 &  18.88 &  -0.57 &  21.708 &  35.38 &  -0.51 &         &        &        \\
\hline
\end{tabular}
\end{center}
\end{table*}

\begin{table*}[tbp]
\begin{center}
\caption{Coordinates of the centers of the areas with the correlation
coefficients $|{\rm K}|\ge0.5$. The coordinates:
right ascension $\alpha$ in hours and declination $\delta$ in degrees
(for the epoch of 2000.0) are accurate
to about half the pixel size. The size of the pixel side is  $6.75\degr$
}
\begin{tabular}{r|r|r||r|r|r||r|r|r}
\hline
$\alpha^h$&$\delta^\circ$&$k$&$\alpha^h$&$\delta^\circ$&$k$&$\alpha^h$&$\delta^\circ$&$k$\\
\hline
5.258  &  -39.62 &  -0.52 & 21.660 &  -17.12 &   0.52 & 12.155 &   31.62 &   0.73  \\
10.516 &  -39.62 &  -0.54 & 1.812  &   -6.62 &   0.54 & 17.406 &   43.62 &   0.66  \\
12.269 &  -39.62 &  -0.50 & 14.617 &    9.88 &  -0.60 & 7.039  &   54.88 &   0.52  \\
10.904 &  -34.38 &  -0.55 & 0.478  &   19.62 &  -0.58 &        &         &         \\
0.942  &  -17.12 &   0.68 & 3.822  &   19.62 &  -0.57 &        &         &         \\
\hline
\end{tabular}
\end{center}
\end{table*}

\begin{table*}[tbp]
\begin{center}
\caption{
The number of events where the correlation coefficient exceeds ${\rm K}=0.5$
in the ``ILC--NVSS'' scans and
in the average coefficient inferred from simulated data. The estimates
are made in the scans with the same declination}
\begin{tabular}{c|c|c}
\hline
Angular scale, &   Number  of         & Average number of         \\
degrees     &   ``ILC--NVSS'' events & ``model--NVSS'' events  \\
\hline
  0.75   &    53       &            51.2$\pm$12.1            \\
  3.0    &    34       &            22.1$\pm$6.0             \\
  4.5    &    10       &            20.0$\pm$6.1             \\
  6.75   &    13       &            15.3$\pm$5.4             \\
\hline
\end{tabular}
\end{center}
\end{table*}

To produce sections of the  ILC map we use the {\it mapcut} procedure of
the GLESP package
\cite{glesp2:Verkhodanov_n}, which allows sections of the map to be made
at the given declination with the
given step in right ascension. Before performing this operation,
we generated maps on the celestial sphere using the
coefficients of spherical harmonics from the $a_{\ell m}$ set with
the given resolution by applying
the {\it cl2map} procedure from the same package with the function of
map synthesis:
\begin{eqnarray}
\Delta T(\theta,\phi)=\sum_\ell^{\ell_{max}}\sum_{m=1}^\ell \hspace{2cm}\nonumber\\\left(a_{\ell,m}
Y_{\ell,m}(\theta,\phi)+a_{\ell,-m}Y_{\ell,-m}(\theta,\phi)\right)\,,
\end{eqnarray}
\noindent
where $Y_{\ell,m}$ are the spherical functions; $\ell$ is the number
of the multipole; $m$ is the mode of the
multipole  $\ell$; $(\theta,\phi)$ are the polar coordinates,
and $a_{\ell,m}$ are the coefficients of spherical
harmonics satisfying the following condition:
\begin{eqnarray}
Y_{\ell,-m}(\theta,\phi) =(-1)^mY^*_{\ell,m}(\theta,\phi),\nonumber\\
a_{\ell,m}=(-1)^m a^{*}_{\ell,-m}\,.
\end{eqnarray}

\section{CORRELATION ANALYSIS}

The adopted approach is based on a two-steps one-dimensional correlation
analysis between the  $M$ scans
of the NVSS and the sections of the  ILC map performed in order to search
for correlation features on various
scales.

At the first stage we selected correlated scans of these two maps and
identified the cases where the level of
correlations was higher than the average level estimated from simulated
records (Fig. 1). Although such a low
correlation level makes the concept of correlation between the scans
sounds problematic, we nevertheless
took into account the fact that small-scale correlations, if present,
must  show up above the Gaussian noise and
rise the correlation level ${\rm K}$ above the statistically expected value.

\begin{figure*}[tbp]
\centerline{
\vbox{
\psfig{figure=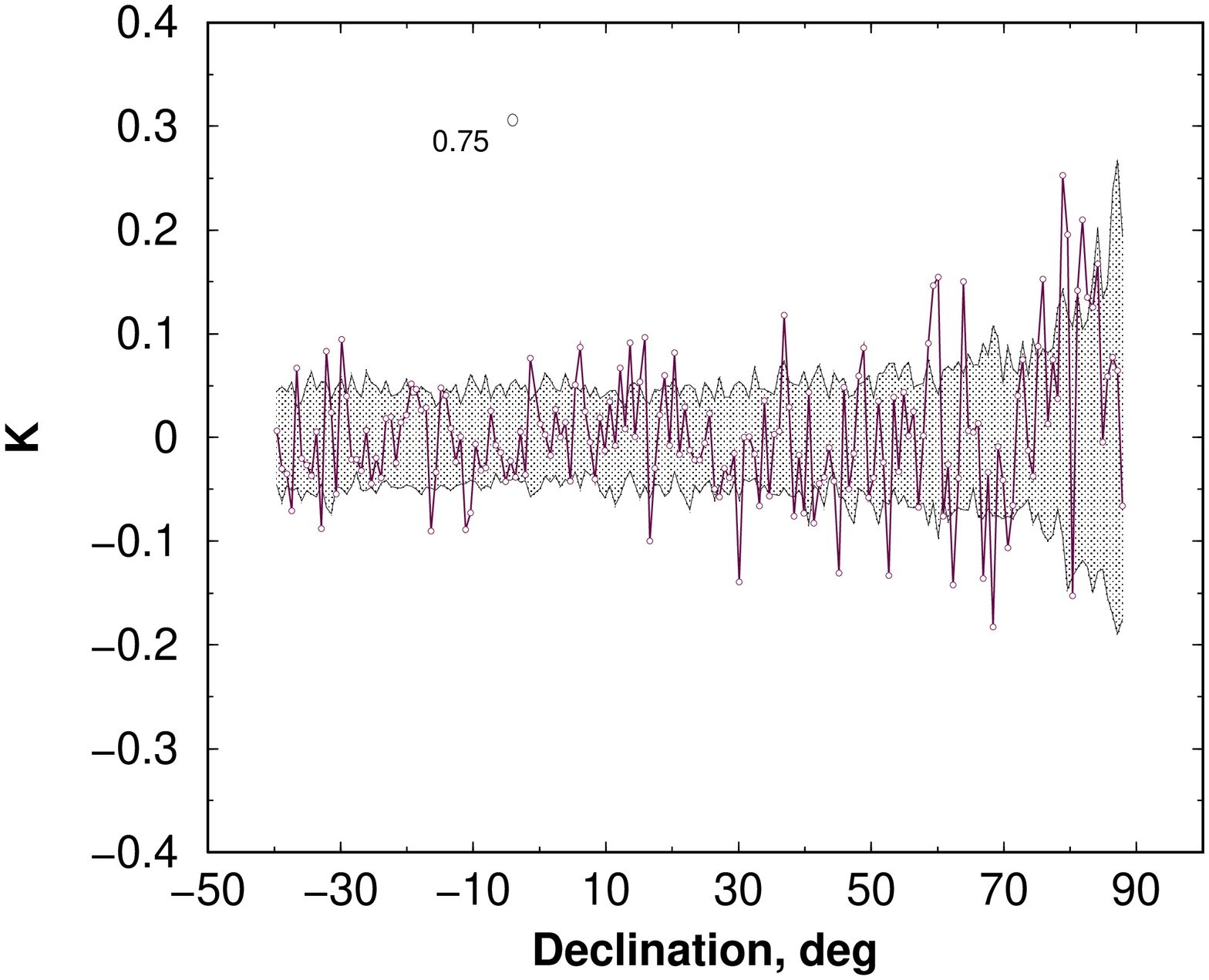,width=7cm,height=3.5cm,angle=-90}
\psfig{figure=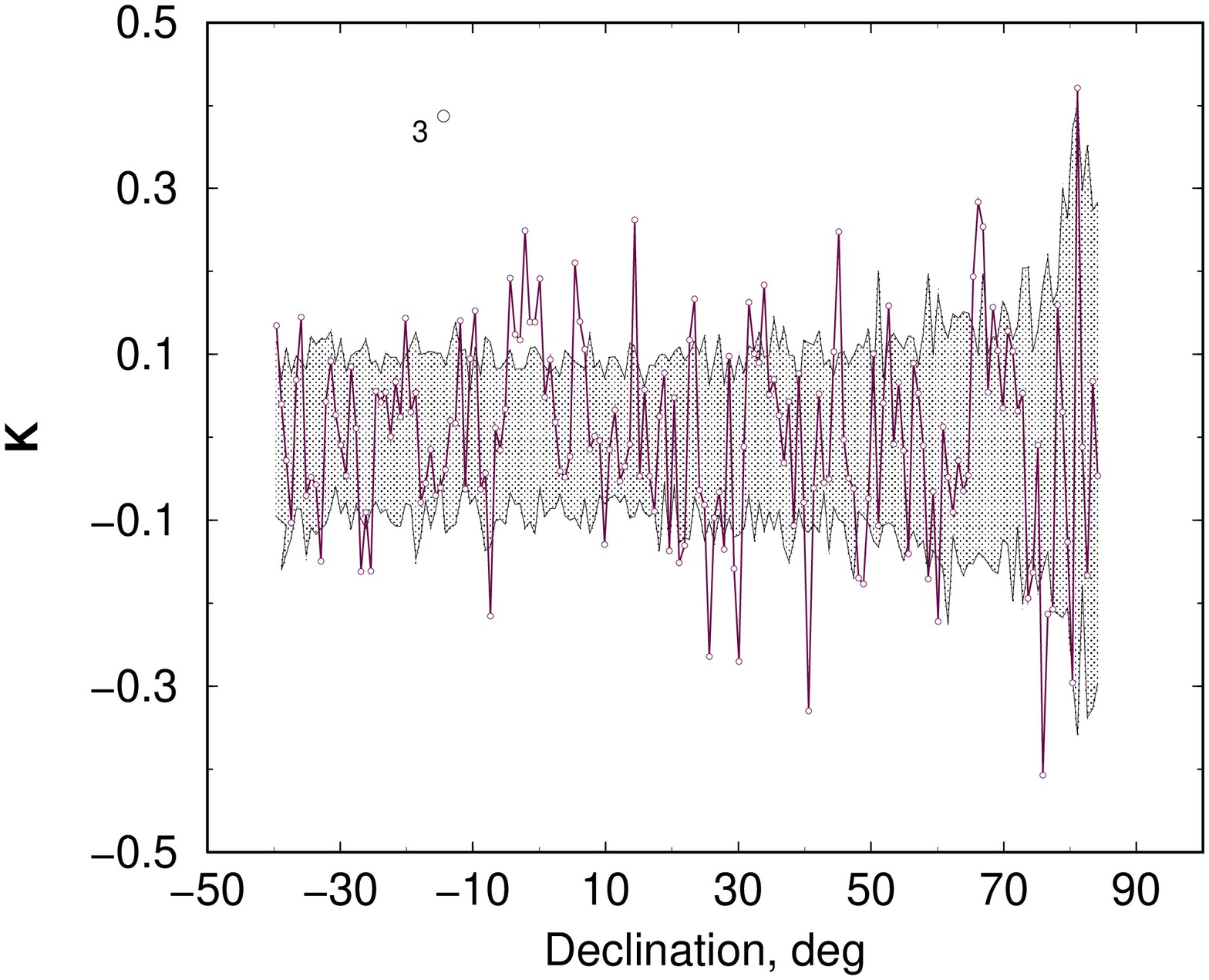,width=7cm,height=4cm,angle=-90}
\psfig{figure=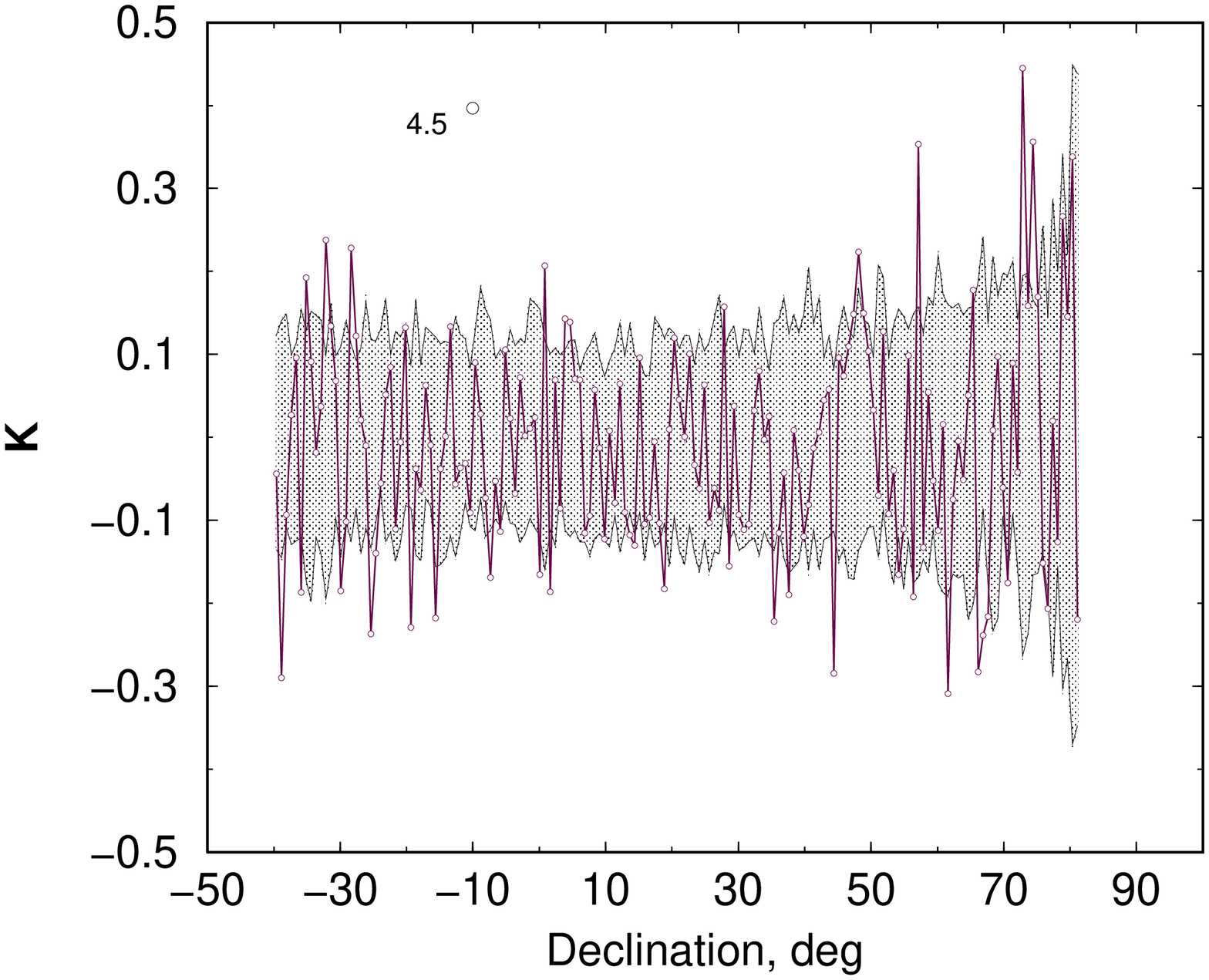,width=7cm,height=4cm,angle=-90}
\psfig{figure=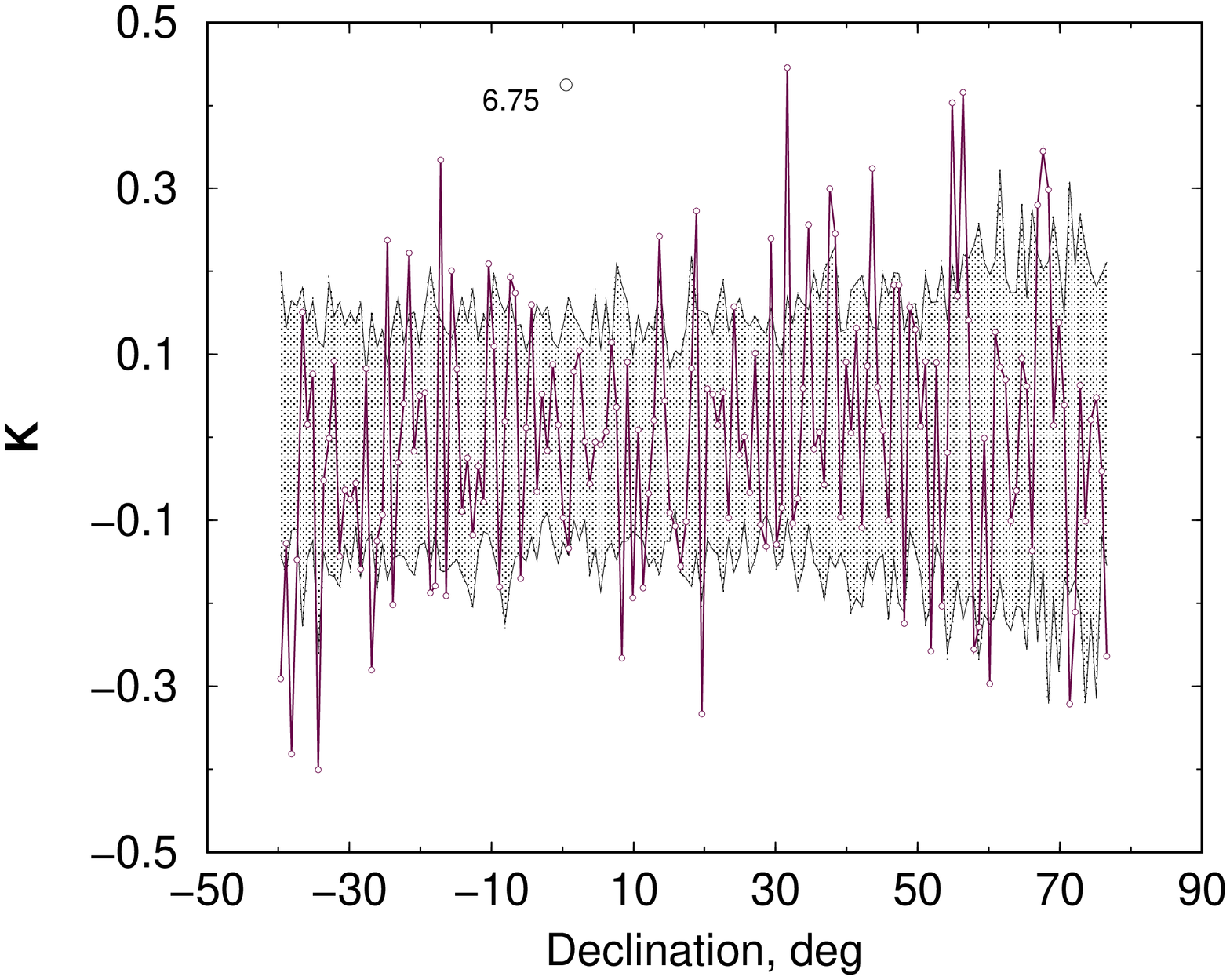,width=7cm,height=4cm,angle=-90}
}}
\caption{
Correlation coefficients for different angular scales as a function of
declination $\delta$.
The correlation scales (0.75, 3, 4.5, and 6.75 degrees) are indicated
in the figures.
Gray shade shows the allowable level of cross correlations as computed
based on the data for
50 simulated maps for  $\Lambda$CDM cosmology.
}
\end{figure*}

\begin{figure*}[tbp]
\centerline{
\psfig{figure=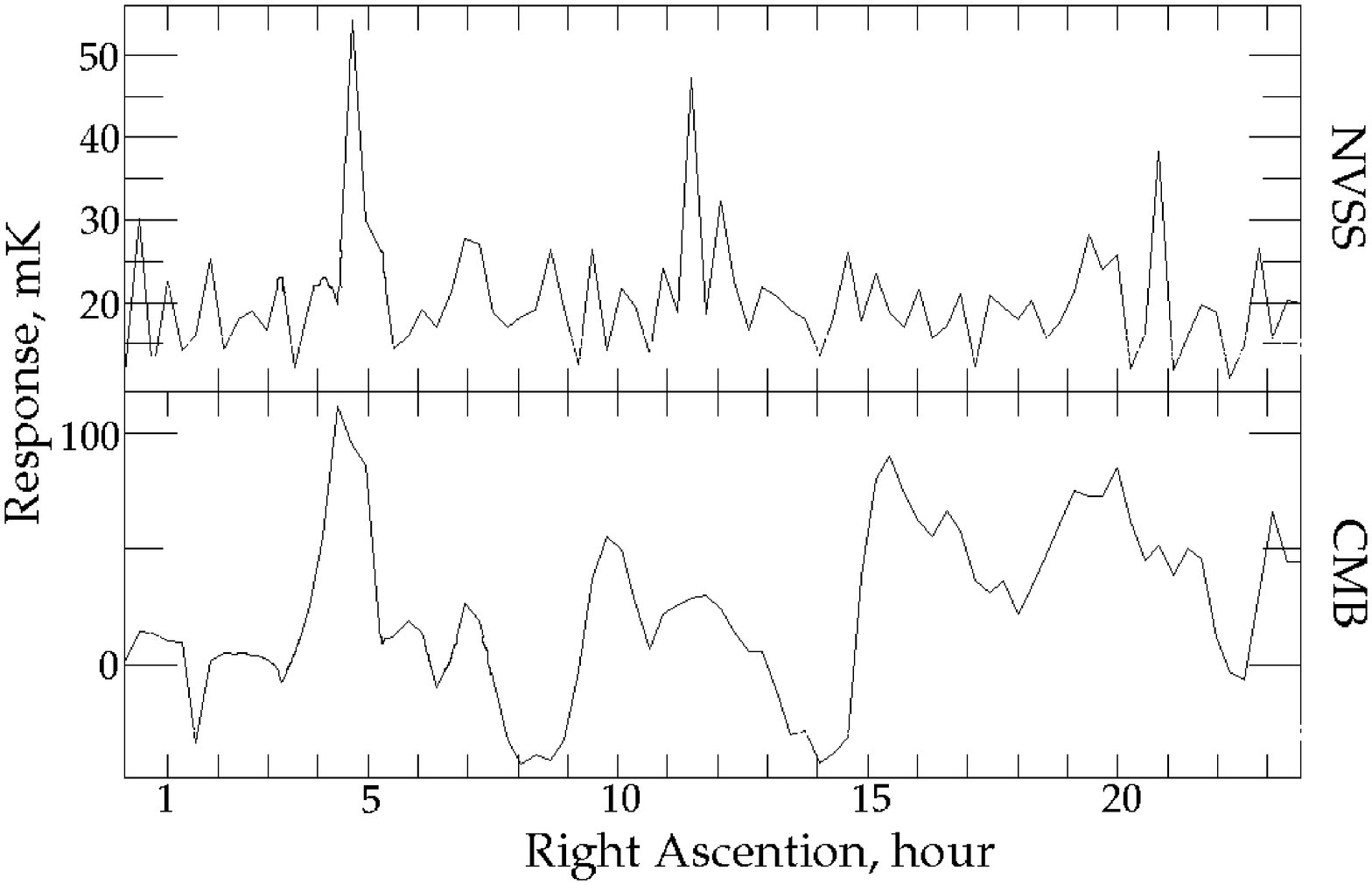,width=9cm,angle=-90}
}
\caption{Examples of  NVSS (upper plot) and CMB (WMAP ILC) (lower plot)
sections at 4.5-degree intervals,
where the correlation coefficient $|{\rm K}|\ge0.5$.}
\end{figure*}

We use standard method to compute the correlation coefficients ${\rm K}_t$
for one-dimensional sections:
\begin{eqnarray}
 {\rm K}_t=\frac{{\rm cov}(x_{ILC,t},\, x_{NVSS, t})}{\sigma_{ILC,t}\sigma_{NVSS,t}} \hspace{2cm}\nonumber\\
 = \frac{\sum\limits_{i=1}^{n}(x_{i,ILC,t}-\overline{x_{ILC,t}})
    (x_{i,NVSS,t}-\overline{x_{NVSS,t}})}
     {\sigma_{ILC,t}\sigma_{NVSS,t}}\,,
\end{eqnarray}
where $x_{i,ILC,t}$ is  $i$th element of the one-dimensional section of
the ILC map written in the form of
vector $x_{ILC,t}$ for the given scale or coordinate interval $t$;
$x_{i,NVSS,t}$ is the similar quantity,
where vector $x_{NVSS,t}$ is based not on a section of the  ILC map,
but on the $M$ scan of NVSS computed
by formula (1); $\overline{x_{ILC,t}}$ and $\overline{x_{fgd,t}}$ are
the data vectors for the sections
of the ILC maps and $M$ sections of NVSS respectively and $\sigma_{ILC,t}$
and $\sigma_{NVSS,t}$ are
the corresponding variances.

The second stage consisted in analyzing the properties of the selected
sections in order to search for
intervals with correlation coefficients $|{\rm K}|\ge0.5$.
Let us point out from the start that if the
procedure involves few points, correlations between the  $M$-sections and
NVSS data  can always
be found even in two random records. Our simulation of pure Gaussian
noise using procedure  {\it fpatt}
of the \mbox{FáDPS reduction system
\cite{fadps:Verkhodanov_n,fadps2:Verkhodanov_n}} showed that the
optimum number of counts such that the correlation coeffciient for
two random records is $|{\rm K}|<0.5$
should be no less than 26. We adopted this number of counts as our
minimum interval in pixels to search
for significant correlations ($|{\rm K}|\ge0.5$) in the records selected
at the first stage. Figure~2
shows an example of such a record, which contains an  interval with
$|{\rm K}|\ge0.5$.

\section{RESULTS}

Due to the insufficient number of pixels in scans we did not use the data
on for angular scales 9.75$\degr$
and thus excluded regions with the sizes on the order of that of
the Cold Spot. We analyze the sections and
maps, constructed for the angular scales of 0.75, 3, 4.5, and 6.75$\degr$.
Tables~\,1--4 list the central coordinates of the intervals and
the correlation coefficients.

The number of events with ${\rm K}\le-0.5$ (34 of 53, 19 of 34, 7 of 10,
and 7 of 13 for the angular scales of
0.75, 3, 4.5, and 6.75$\degr$ respectively) exceeds a bit the number of
events with ${\rm K}\ge0.5$. This result
could be explained by invoking the Sachs--Wolfe effect however,
our subsequent statistical simulation
shows that such deviations are within the limits of Gaussian scatter.

To test the statistical significance of the results with
the Monte Carlo technique we used GLESP package
\cite{glesp:Verkhodanov_n} to generate 50 simulated random maps of
microwave background for different angular scales
with a power spectrum corresponding to  $\Lambda$CDM cosmology.
We constructed our maps with pixelization on
different angular scales similar to the pixelization of the maps studied.
On each map we identified the scans at
the declinations listed in the previous table and computed the number of
events with the correlation
coefficient $|{\rm K}|\ge0.5$. We list the computed number of events
in Table~\,5.

The results listed in Table~\,5 indicate that no statistically significant
deviations can be found
between the NVSS map---ILC CMB map correlation and the NVSS map---random CMB
maps correlation on angular
scales of (1--7)$\degr$. However, we identified the domains
listed in the catalog (Tables~1--4), where the behavior of $M$-scans
resembles that of CMB. It is not improbable
that the correlation between the positions of some spots from the catalog
of celestial areas may be due to a real
physical effect, although we would rather conclude that most of
the correlated (anticorrelated) spots (or perhaps even
all spots) can be explained as simple statistical coincidences.

\section{CONCLUSIONS}

We performed a correlation analysis of the CMB maps and of the modified
NVSS map smoothed at different angular scales
using the method of one-dimensional sections. We constructed the modified
NVSS map using the distribution of
the mean squared flux of sources within an area of given size centered
on the pixel specified by the selected
area. We used the data obtained as a result of our search for correlations
to catalog the regions with
$|{\rm K}|\ge0.5$ at various angular scales. We simulated the CMB maps
using the Monte Carlo method and the angular
power spectrum of the $\Lambda$CDM cosmology as our input parameter.
We performed a similar correlation analysis
for 50 generated CMB models. We found that the statistical properties
of the correlations on the angular scales
studied (0.75, 3, 4.5, and 6.75$\degr$) does not differ from those of
random maps. This result leads us to
conclude that the correlations found may be due to statistical coincidence.
However, it is of interest to perform
two-dimensional correlation pixelization (mapping) of the sky using catalogs
for other wavelength intervals. We plan
to perform such an analysis in our next paper. We are going to,
in particular, to look for the Sachs--Wolfe effect.

\noindent
{\small
{\bf Acknowledgments}.
We are grateful to  NASA for making available their NASA Legacy Archive,
from where we adopted  \wmap data.
We are also grateful to the authors of  HEALPix for the use of their
\footnote{\tt http://www.eso.org/science/healpix/} package
\cite{healpix:Verkhodanov_n}, which we used to transform \wmap maps
into the coefficients $a_{\ell m}$.
In this paper we used the
\glesp \footnote{\tt http://www.glesp.nbi.dk} package
\cite{glesp:Verkhodanov_n,glesp2:Verkhodanov_n}
for subsequent analysis of the CMB data on the sphere and the FADPS system
\footnote{\tt http://sed.sao.ru/$\sim$vo/fadps\_e.html} for the reduction
of one-dimensional  data
\cite{fadps:Verkhodanov_n,fadps2:Verkhodanov_n}.
This work was supported in part by the grant  ``Leading scientific schools
of Russia''.
O.V.V. acknowledges partial support from the Russian Foundation Basic
Research for partial support of the work
(grant no.~\,08-02-00159) and Russian Foundation for Domestic Science
Support.
}

\end{document}